\documentclass[aps,pre,twocolumn,superscriptaddress,amsmath,amsfonts,showpacs]{revtex4}
\usepackage{amsmath}
\usepackage{amssymb}
\usepackage{graphicx}
\usepackage{subfigure}
\usepackage{hyperref}
\hypersetup{bookmarks,
            colorlinks,
            citecolor=red,
            filecolor=blue,
            urlcolor=blue,
            pdfpagemode=None}

\begin{document}

\title{Optimal operating points of oscillators using nonlinear resonators}%
\author{Eyal Kenig}
\email[Corresponding author:\ ]{eyalk@caltech.edu}
\affiliation{Kavli Nanoscience Institute and Condensed Matter Physics, California Institute of Technology, MC 149-33, Pasadena, California 91125, USA}
\author{M. C. Cross}
\affiliation{Kavli Nanoscience Institute and Condensed Matter Physics, California Institute of Technology, MC 149-33, Pasadena, California 91125, USA}
\author{L. G. Villanueva}
\affiliation{Kavli Nanoscience Institute and Condensed Matter Physics, California Institute of Technology, MC 149-33, Pasadena, California 91125, USA}
\author{R. B. Karabalin}
\affiliation{Kavli Nanoscience Institute and Condensed Matter Physics, California Institute of Technology, MC 149-33, Pasadena, California 91125, USA}
\author{M. H. Matheny}
\affiliation{Kavli Nanoscience Institute and Condensed Matter Physics, California Institute of Technology, MC 149-33, Pasadena, California 91125, USA}
\author{Ron Lifshitz}
\affiliation{Raymond and Beverly Sackler School of Physics and Astronomy, Tel Aviv University, 69978 Tel Aviv, Israel}

\author{M. L. Roukes}
\affiliation{Kavli Nanoscience Institute and Condensed Matter Physics, California Institute of Technology, MC 149-33, Pasadena, California 91125, USA}

\begin{abstract}
We demonstrate an analytical method for calculating the phase sensitivity of a class of oscillators whose phase does not affect the time evolution of the other dynamic variables. We show that such oscillators posses the possibility for complete phase noise elimination. We apply the method to a feedback oscillator which employs a high $Q$ weakly nonlinear resonator, and provide explicit parameter values for which the feedback phase noise is completely eliminated, and others for which there is no amplitude-phase noise conversion. We then establish an operational mode of the oscillator which optimizes its performance by diminishing the feedback noise in both quadratures, thermal noise, and quality factor fluctuations. We also study the spectrum of the oscillator, and provide specific results for the case of $1/f$ noise sources.
\end{abstract}
\date{\today}


\maketitle


\section{Introduction}
The importance of oscillators is widely expressed in both natural and engineered systems. Such devices, generating a periodic signal at an inherent frequency, are often developed to serve as  highly accurate time or frequency references~\cite{vig}. Since the ideal self-sustained oscillator is mathematically described as a periodic solution of a set of autonomous differential equations, its steady state dynamics can be expressed in terms of the phase variable tracing the motion along a limit cycle. This phase is highly sensitive to the stochastic noise present in the physical system, as the appearance of periodicity without an external time reference implies the freedom to drift along the phase direction. The noise induced phase drift causes an unwanted broadening of the oscillator's spectral peaks, which ideally would have been perfectly discrete~\cite{lax67}. Therefore, the phase sensitivity to noise, which quantifies the oscillator's performance under the influence of an arbitrary noise spectrum, should be as small as possible.

In this paper we study oscillators that are well described by equations of motion that are independent of the oscillator phase variable, so that the phase variable does not affect the time evolution of the dynamical system. Common examples that fall in this class are oscillators near the Hopf bifurcation generating the periodic motion, and the standard reduced envelope description of oscillators based on a high $Q$ linear or weakly nonlinear resonator, with the oscillations sustained by energy injection either through an active feedback loop~\cite{lax67,YurkePra,greywall,pfo}, or by nondegenerate parametric excitation~\cite{kenig12}. An important consequence of this feature is that the limit cycle has circular shape, and the complete stochastic problem can be resolved analytically as we show in Sec.~\ref{sec1}. In the absence of this symmetry, the stochastic problem can only be treated numerically, using methods such as those proposed by Demir~\cite{Demir00,Demir02,isochrone,suvak}. Besides this computational advantage, a remarkable physical property of this class of oscillators is the possibility for complete noise elimination by tuning a single parameter, which cannot be achieved in the general case.

In Sec.~\ref{app}, as an example we calculate the phase sensitivity to various noise sources of the standard feedback oscillator based on a high $Q$ weakly nonlinear resonator. It has previously been shown  that for a saturated feedback amplitude this oscillator is insensitive to the feedback phase noise at the critical Duffing point~\cite{YurkePra,greywall}. We show that this property is a particular case of parameter noise elimination, and provide explicit expressions for two feedback loop phase-shift values as a function of the amplifier saturation level, for which feedback phase noise is completely eliminated, and an additional value that eliminates the amplitude-phase noise conversion by the nonlinear resonator. We show that in the high saturation limit, one of the operating points that eliminates the feedback phase noise also eliminates noise in the feedback amplitude, additive thermal noise, and quality factor fluctuations. Operating at this point at the high saturation limit is thus an optimal operational state of the oscillator, as it was recently demonstrated using an oscillator based on a nanomechanical resonator~\cite{guillermo}.

We also use the simple analysis possible for phase-symmetric oscillators to derive complete expressions for the spectrum of the oscillator for both white and colored noise sources. For the special case of $1/f$ noise sources, often encountered in physical implementations, we show that the near carrier spectrum is well approximated by a Lorentzian for weak noise, and by a Gaussian for stronger noise. In both cases the tails of the spectral peaks away from the carrier frequency $\omega_{0}$ fall off as $(\omega-\omega_{0})^{-3}$, as is known from previous results.

\section{Noise Projection method}\label{sec1}
We start by describing an analytical method for calculating the stochastic phase sensitivity of a class of oscillators described by the $N$ equations of motion
\begin{equation}\label{}
    \dot{\textbf{X}}=\textbf{f}(X_1,...,X_{N-1}),
\end{equation}
where $\textbf{X}^T=(X_1,...,X_{N-1},\Phi)$ is the vector of dynamical variables, and $\textbf{f}^T=(f_1,...,f_N)$. Note that $\textbf{f}$ is not a function of the phase variable $\Phi$. For the oscillating state, the first $N-1$ equations have a fixed point $(X_{1,0},...,X_{N-1,0})$, and the phase continuously rotates at the rate $\omega_0=f_N(X_{1,0},...,X_{N-1,0})$.
This is the deterministic description of a class of oscillators for which the phase variable $\Phi$ does not affect the time evolution of the dynamical system. Once we add a particular small random noise defined by the stochastic scalar $\Xi(t)$ to this deterministic description and linearize the equations near the fixed point describing the limit cycle, the dynamics of the small perturbations $\textbf{x}^T=(x_1,...,x_{N-1},\phi)$ are described by the equations
\begin{equation}\label{lin}
   \dot{\textbf{x}}=\textbf{J}\textbf{x}+\textbf{v}_{\textmd{noise}}\Xi,
\end{equation}
with $\textbf{v}^T_{\textmd{noise}}(X_{1,0},...,X_{N-1,0})=(v_{\textmd{noise},1},...,v_{\textmd{noise},N})$ describing how the noise couples to the dynamical variables, and the Jacobian
\begin{equation}\label{}
 \textbf{J}=\left.\begin{pmatrix}
 \frac{\partial f_1}{\partial X_1} & ... & \frac{\partial f_1}{\partial X_{N-1}} & 0 \\ \vdots & \vdots & \vdots & \vdots \\ \frac{\partial f_N}{\partial X_1} & ... &\frac{\partial f_N}{\partial X_{N-1}}& 0
 \end{pmatrix}\right|_{X_{1,0},...,X_{N-1,0}}.
\end{equation}

Since the oscillator spectral peak is typically much narrower than the corresponding line width of the driven resonator, which is determined by the decay rates that are the negative eigenvalues $\lambda_i$ of the Jacobian, we are mostly interested in frequency offsets satisfying $|\omega-\omega_0|\ll|\lambda_i|$. This requirement is equivalent to taking the time derivatives of the first $N-1$ variables of Eq.~(\ref{lin}) to be zero.
If we then multiply Eq.~(\ref{lin}) from the left by some arbitrary vector $\textbf{v}^T=(v_1,...v_{N-1},1)$, we obtain a scalar equation for the time evolution of the phase perturbation
\begin{equation}\label{phidot}
   \dot{\phi}=\textbf{v}^T\textbf{J}\textbf{x}+\textbf{v}^T\textbf{v}_{\textmd{noise}}\Xi.
\end{equation}
Now taking $\textbf{v}=\textbf{v}_\bot$, where $\textbf{v}_\bot$ is the eigenvector of the transposed Jacobian that corresponds to the zero eigenvalue, for which $\textbf{J}^T\textbf{v}_\bot=0$, eliminates the first term on the right hand side of Eq.\ (\ref{phidot}), and we are left with
\begin{equation}\label{proj}
   \dot{\phi}(t)=\textbf{v}_\bot^T\textbf{v}_{\textmd{noise}}\,\Xi(t).
\end{equation}

Thus we find the result that the noise driven phase evolution is determined by the scalar product of the noise vector and the zero mode of the transposed Jacobian. This scalar product quantifies the phase sensitivity to a particular noise vector. The geometrical interpretation of this result is that in the linear approximation in the rotating frame, each point on the limit cycle $\Phi_0$ is associated with a hyperplane $H_{\Phi_0}$ spanned by the eigenvectors of the Jacobian corresponding to negative eigenvalues, and all points in this hyperplane asymptote to $\Phi_0$. Therefore, any perturbations in this hyperplane have no long term effect on the phase. The zero-eigenvalue eigenvector of the transposed Jacobian is orthogonal to this hyperplane, and thus projecting the noise vector onto this vector accounts for the noise induced phase drift. Under the full nonlinear flow, the set $I_{\Phi_0}$ in the stable manifold of the limit cycle that asymptotes to $\Phi_0$ is called the \textit{isochrone}~\cite{isochrone,moehlis,winfree},  and $H_{\Phi_0}$ is its linear approximation at the vicinity of the limit cycle.

In Sec.~\ref{spec} we use Eq.~(\ref{proj}) and express the oscillator spectrum as a function of the noise spectrum. However, the phase noise will be proportional to the phase sensitivity $(\textbf{v}_\bot^T\textbf{v}_{\textmd{noise}})^2$ regardless of the particular spectral shape. This quantity thus provides an experimentally controlled measure of the oscillator performance, under the influence of an arbitrary noise spectrum.
The only change necessary for applying Eq.~(\ref{proj}) to a general limit cycle, is to replace the two vectors with their time dependent counterparts $\textbf{v}_\bot^T\textbf{v}_{\textmd{noise}}\rightarrow\textbf{v}_\bot(\omega_0t+\phi(t))^T\textbf{v}_{\textmd{noise}}(\omega_0t+\phi(t))$, as derived by many authors \cite{Demir00,isochrone,nakao}. The fact that in the present case these vectors are constant leads to a particularly simple derivation and analysis, and grants the system the possibility of complete noise elimination by tuning a single parameter to a point at which the scalar product is zero, as we show next.

\subsection{Parameter Noise}\label{sec: Parameter Noise}
Let us consider noise which derives from fluctuations in some parameter $p$ of the equations so that $\textbf{v}^T_{\textmd{noise}}=\left(\partial f_1/\partial p,...,\partial f_N/\partial p\right)$.
The fixed points satisfy the following equations
\begin{eqnarray}\label{}
  0 &=& f_1(X_{1,0},...,X_{N-1,0};p), \\
  &\vdots&\nonumber\\
  0 &=& f_{N-1}(X_{1,0},...,X_{N-1,0};p),\nonumber\\
  \omega_0 &=&   f_N(X_{1,0},...,X_{N-1,0};p)\nonumber.
\end{eqnarray}

First, we find the effect of a \emph{change} in the parameter $p$ on the mean oscillator frequency by differentiating these equations and multipling from the left by $\textbf{v}_\bot^T$ to find
\begin{equation}\label{dOmega}
    d\omega_0=\textbf{v}_\bot^T\textbf{J}\textbf{dX}_0+\textbf{v}_\bot^T\textbf{v}_{\textmd{noise}}dp,
\end{equation}
with the \emph{same} vector $\textbf{v}_{\textmd{noise}}$ appearing as in the stochastic problem for noise in the parameter $p$.
Since the first term on the right hand side of Eq.\ (\ref{dOmega}) is zero we get
\begin{equation}\label{pnoise}
    \frac{d\omega_0}{dp}=\textbf{v}_\bot^T\textbf{v}_{\textmd{noise}}.
\end{equation}
Now returning to the stochastic problem, we see that the measure which quantifies the phase wandering as a result of parameter noise, is just the derivative of the frequency with respect to the parameter containing the noise. This implies that for the oscillators under consideration, phase fluctuations due to noise in a particular parameter can be completely eliminated by tuning the system to the point at which $d\omega_0/dp=0$.

\subsection{Amplitude-phase noise conversion}\label{sec: Amplitude-phase noise conversion}
Other noise sources, such as additive thermal noise, will not in general be equivalent to a parameter fluctuation, and typically it will not be possible to eliminate the resulting phase noise completely. To see how to reduce the effects of such a noise source, we write the stochastic equation for the phase variable in Eq.~(\ref{lin}) explicitly
\begin{align}\label{ampPhaseGeneral}
    \dot{\phi}=\sum_{n=1}^{N-1}&\frac{\partial f_N(X_{1,0},...,X_{N-1,0})}{\partial X_n}x_n\nonumber\\
    &+v_{\textmd{noise},N}(X_{1,0},...,X_{N-1,0})\Xi.
\end{align}
While the last term originates from the noise force acting directly on the phase variable, the first $N-1$ terms account for the phase drift due to noise conversion from other fluctuating variables $x_n$ to the phase.
For oscillators based on a high-$Q$ resonator, amplitude-phase noise conversion in a simple two-variable description is often depicted as limiting the operating amplitude, since this source of phase noise   increases when the resonator is driven into the nonlinear regime. However, we now see that the conversion of noise from the variable $x_n$ to the phase is completely eliminated by operating at $\partial\omega_0/\partial X_n=0$. Note that this is a partial derivative, in contrast with the full derivative that appears in Eq.~(\ref{pnoise}).

\section{Feedback oscillator based on a nonlinear resonator}\label{app}
Precise time and frequency references are typically built around high $Q$ resonators, since high-$Q$ reduces phase noise in the linear regime \cite{Leeson66}. In the high $Q$ limit, the standard feedback oscillator is effectively described by a complex amplitude equation that captures the slow dynamics near the high natural frequency $\bar{\omega}_0$ of its weakly nonlinear resonating element~\cite{lax67,YurkePra, LCreview,pfo}
\begin{equation}\label{ampEq}
    \frac{dA}{dT}=\left(-\frac{1}{2}+i\frac{3}{8}|A|^2\right)A-i\frac{s}{2}e^{i\Phi}e^{i\Delta},
\end{equation}
where $T=t\bar{\omega}_0/Q$  is a slow time that scale, $A=ae^{i\Phi}$ is the complex amplitude, $\Delta$ is the phase shift of the feedback, and $s$ is the feedback level, which is assumed to be the output of a saturated amplifier and so is independent of the magnitude $|A|$ of oscillation \footnote{The analysis is easily extended the case of a nonsaturated amplifier.}. Separating Eq.~(\ref{ampEq}) into real and imaginary parts gives the equations
\begin{eqnarray}\label{ampeq}
  \frac{da}{dT}&=&-\frac{ a}{2}+\frac{s}{2}\sin\Delta=f_a(a),\\
  \frac{d\Phi}{dT} &=& \frac{3}{8}a^2-\frac{s}{2}\frac{\cos\Delta}{a}=f_{\Phi}(a)\nonumber.
\end{eqnarray}

In correspondence with the general description of the previous section, the first equation has the fixed point $a_0=s\sin\Delta$, the oscillation frequency is $\Omega_0=f_{\Phi}(a_0)$, and the projection vector given by the zero mode of the transposed Jacobian is
\begin{equation}\label{zeroMode}
    \textbf{v}_\bot^T=\left(-\frac{\left(\frac{\partial f_{\Phi}}{\partial a}\right)}{\left(\frac{\partial f_a}{\partial a}\right)},1\right)=\left(\frac{\cos\Delta}{s\sin^2\Delta}+\frac{3}{2}s\sin\Delta,1\right).
\end{equation}
We can now analyze the oscillator phase noise due to various anticipated noise sources.

\subsection{Feedback noise}\label{sec: Feedback noise}
Noise in the saturated feedback signal is expressed in the complex amplitude description by two noise vectors, one representing fluctuations in the feedback amplitude, and the other representing fluctuations in the feedback phase. Since the noise in the feedback phase is more dominant for a saturated feedback \cite{YurkePra,greywall}, as a first example of our formalism we calculate the phase sensitivity to this noise source, which is expressed by fluctuations in the parameter $\Delta$. The corresponding noise vector is $\textbf{v}_{\Delta}^T=(\partial f_a/\partial \Delta,\partial f_{\Phi}/\partial\Delta)$. Using the explicit form of $\textbf{v}_{\Delta},\textbf{v}_\bot$ or the expression derived in \S\ref{sec: Parameter Noise} gives
\begin{equation}\label{ampPhaseNoise}
    \textbf{v}_\bot^T\textbf{v}_{\Delta}=\frac{d\Omega_0}{d\Delta}=\frac{3 s^2\sin\Delta\cos\Delta}{4}+\frac{1}{2\sin^2\Delta}.
    \end{equation}
At two phase shift values $\Delta_{1,2}$, this expression is zero and the oscillator is insensitive to the phase noise of the feedback signal. The explicit values of $\Delta_{1,2}$ are given in Eq.~(\ref{delta12}) in the Appendix.
These points bifurcate at the critical Duffing point for which $\Delta_1=\Delta_2=2\pi/3$. As we already noted, the ability to eliminate the feedback phase noise at the critical Duffing point has been previously established~\cite{YurkePra,greywall}.

The second contribution of the feedback noise to the phase drift comes from noise in the feedback level. This noise is expressed by fluctuations in the parameter $s$, for which the noise sensitivity is
\begin{equation}\label{ampAmpNoise}
    \textbf{v}_\bot^T\textbf{v}_{s}=\frac{d\Omega_0}{ds}=\frac{3s\sin^2\Delta}{4}.
\end{equation}
This noise cannot be eliminated in the same manner as feedback phase noise, since the oscillator amplitude is zero at $\Delta=n\pi$, however we show in \S\ref{sec: combined} how it can be significantly reduced.

\subsection{Amplitude-phase noise conversion}
\begin{figure}[]
  \includegraphics[width=0.95\columnwidth]{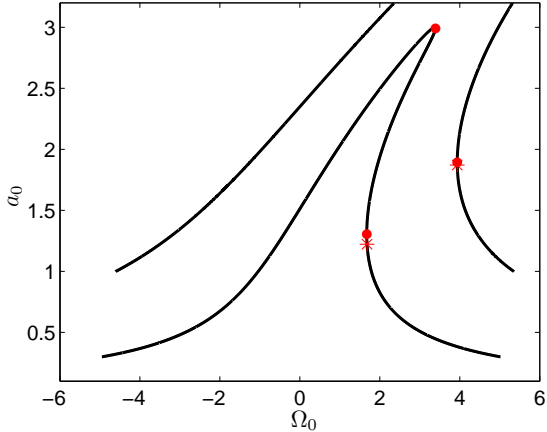}
  \caption{\label{fig1}(Color online) The amplitude vs.~frequency curve of the oscillator described by Eq.\ (\ref{ampEq}). The low and high amplitude red dots are $\Delta_{1,2}$, respectively, for which $d\Omega_0/d\Delta=d\Omega_0/da=0$. The star is $\Delta_{\textmd{a}\mid\phi}$ for which $\partial\Omega_0/\partial a=0$. The bottom curve is for $s=3$, and the top one is for $s=10$, indicating that $\Delta_1$ and $\Delta_{\textmd{a}\mid\phi}$ approach each other as saturation level increases.}
\end{figure}

Based on the analysis of \S\ref{sec: Amplitude-phase noise conversion}, complete elimination of amplitude-phase noise conversion is achieved in the present oscillator for
\begin{equation}\label{ampPhase}
    \frac{\partial\Omega_0}{\partial a_0}=\frac{3}{4}a_0+\frac{s}{2}\frac{\cos\Delta}{a_0^2}=\frac{3s}{4}\sin\Delta+\frac{1}{2}\frac{\cos\Delta}{s\sin^2\Delta}=0,
\end{equation}
whose real solution $\Delta_{\textmd{a}\mid\phi}$ is given in Eq.~(\ref{DeltaAmpPhase}) in the Appendix. As the saturation level grows $\Delta_{\textmd{a}\mid\phi}$ approaches $\Delta_1$ as illustrated in Fig.~(\ref{fig1}), which shows these points on the amplitude-frequency curve for the oscillator, which follows the well known Duffing curve.
Note also that from Eq.~(\ref{ampPhaseNoise}),  we have
\begin{equation}\label{conversionAtYurke}
    \frac{\partial\Omega_0(\Delta_1)}{\partial a_0}=-\frac{1}{2s\cos(\Delta_1)},
\end{equation}
so that as the saturation grows, operating at $\Delta_1$ eliminates the amplitude-phase noise conversion as well. On the other hand, at the operating point $\Delta_{\textmd{a}\mid\phi}$ we have
\begin{equation}\label{}
    \frac{d\Omega_0(\Delta_{\textmd{a}\mid\phi})}{d\Delta}=\frac{1}{2},
\end{equation}
and the feedback phase noise is not reduced here.
Thus, although the points $\Delta_{\textmd{a}\mid\phi},\Delta_1$ approach each other as saturation grows, it is better to operate at $\Delta_1$ than at $\Delta_{\textmd{a}\mid\phi}$.

\subsubsection{Thermal noise}
Thermal noise acting on the resonator degrees of freedom is manifested as equal independent sources in the two quadratures. In the complex amplitude representation, this is described by noise vectors for the two sources, $\textbf{v}^T_{\Re}=(1,0)$, and $\textbf{v}^T_{\Im}=(0,1/a)$.
 The total phase sensitivity to thermal noise is given by adding the two noise terms in quadrature
\begin{eqnarray}\label{}
    \textmd{TPS}(\Delta)&=&(\textbf{v}_\bot^T\textbf{v}_{\Im})^2+(\textbf{v}_\bot^T\textbf{v}_{\Re})^2\\
    &=&\frac{1}{s^2\sin^2\Delta}+\left(\frac{6s}{4}\sin\Delta+\frac{\cos\Delta}{s\sin^2\Delta}\right)^2.\nonumber
\end{eqnarray}
Since the first term is the direct phase noise, and the second results from amplitude-phase conversion, the second term satisfies $\textbf{v}_\bot^T\textbf{v}_{\Re}(\Delta_{\textmd{a}\mid\phi})=0$.
However, the total effect of thermal noise is minimal at the nearby phase shift value $\Delta_{\textmd{TPS}}$, which satisfies $d\textmd{TPS}(\Delta_{\textmd{PST}})/d\Delta=0$.

\subsection{Optimal operating point for combined noise sources}\label{sec: combined}

\begin{figure}[]
  \includegraphics[width=0.95\columnwidth]{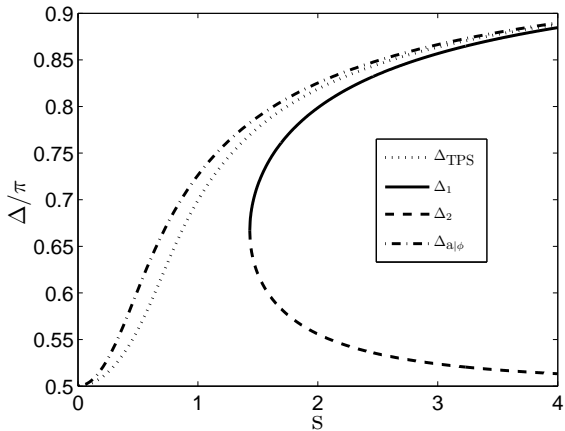}
  \caption{\label{fig2} Special phase shift values as a function of the drive amplitude. $\Delta_{1,2}$, which eliminate the feedback phase noise, bifurcate at the critical Duffing drive amplitude $s=(4/3)^{(5/4)}$. On increasing saturation level, $\Delta_2$ approaches $\pi/2$, and the other three special phase shift values approach $\pi$.}
\end{figure}

The four special phase shift values minimizing the effects of various noise sources on the feedback oscillator are plotted in Fig.~\ref{fig2}, as a function of the other experimentally controlled parameter, the saturation level. Since the real physical oscillator is subjected to multiple, uncorrelated noise sources, we would like to use these results to find its optimal operational state. We know from \S\ref{sec: Feedback noise} that the phase shift values $\Delta_{1,2}$ eliminate the feedback phase noise. Since this is a substantial noise source we would like to eliminate, lets examine the sensitivity to feedback amplitude noise at these points. Using Eq.~(\ref{ampAmpNoise}) and the high saturation limit of Eq.~(\ref{root})  allows us to deduce the limits
\begin{eqnarray}\label{}
   \lim_{s\rightarrow\infty}\textbf{v}_\bot^T\textbf{v}_{s}(\Delta_2)\sim s\rightarrow\infty,\\
    \lim_{s\rightarrow\infty}\textbf{v}_\bot^T\textbf{v}_{s}(\Delta_1)\sim s^{-1/3}\rightarrow0.\nonumber
\end{eqnarray}
Therefore, $\Delta_1$ is a preferable operating point from the feedback noise perspective because for high saturation it eliminates noise in both quadratures of the feedback, while operating at $\Delta_2$ only eliminates the feedback phase noise, with the feedback amplitude noise becoming more important with increasing saturation level.

To consider the effects of thermal noise at $\Delta_1$, we use Eq.~(\ref{conversionAtYurke}), which implies that for large saturation amplitude-phase conversion is diminished at this point, and together with the limit $\lim_{s\rightarrow\infty}a_0(\Delta_1)\sim s^{1/3}$, we derive the results
\begin{eqnarray}
  \lim_{s\rightarrow\infty}\textbf{v}_\bot^T\textbf{v}_{\Re}(\Delta_1)&\sim&s^{-1} \rightarrow 0,\\
 \lim_{s\rightarrow\infty}\textbf{v}_\bot^T\textbf{v}_{\Im}(\Delta_1)&\sim& s^{-1/3}\rightarrow 0.\nonumber
\end{eqnarray}
Thus, in the large saturation limit, operating at $\Delta_1$ also eliminates both components of thermal noise.

\begin{figure}[]
\begin{center}
      \subfigure[]{
   \includegraphics[width=0.95\columnwidth]{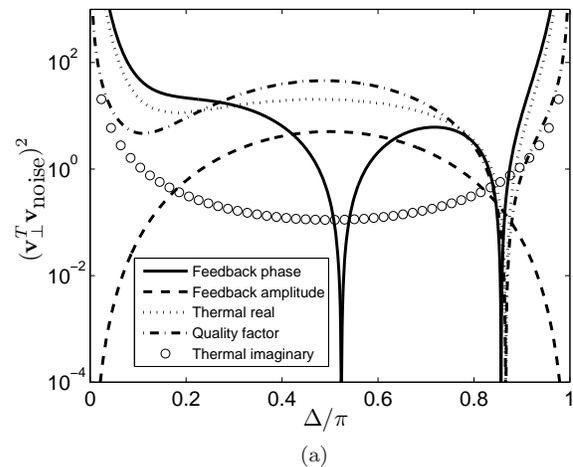}}
  \subfigure[]{
   \includegraphics[width=0.95\columnwidth]{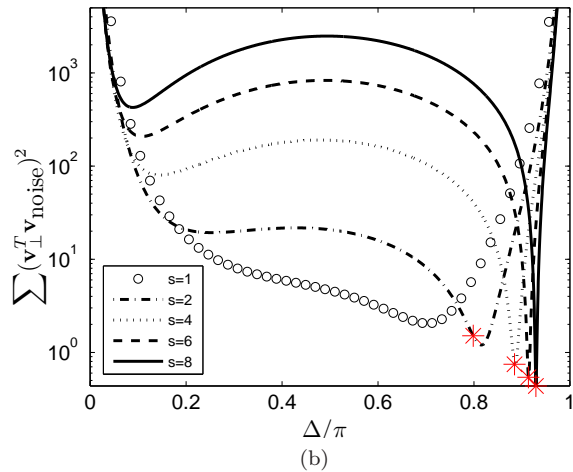}}
   \end{center}
   \caption{\label{fig3}(Color online) Phase sensitivity as a function of the phase shift. (a) Different contributions $(\textbf{v}_\bot^T\textbf{v}_{\textmd{noise}})^2$ to the phase noise, for $s=3$. We see that the feedback phase noise curve is zero at $\Delta_{1,2}$, and that those of the real component of thermal noise and quality factor noise are zero at $\Delta_{\textmd{a}\mid\phi}$, which is slightly to the right of $\Delta_1$
   . (b) The total phase sensitivity, which is a sum of all the curves in (a), for different values of the drive level. We see that as saturation level grows, the optimal operating point approaches $\Delta_1$, noted as red asterisk, and the noise level at this point decreases.}
\end{figure}

Noise in the quality factor is expressed by fluctuations in the linear damping coefficient in Eq.~(\ref{ampEq}). Since we consider the damping to be small, fluctuations with the same relative intensity as other parameters will typically yield a considerably smaller effect on the phase drift.  However, it is instructive to include these fluctuations in the analysis as well. The noise vector representing them is $\textbf{v}^T_{Q}=(-a/2,0)$; it is transferred to phase noise through amplitude-phase conversion and satisfies
\begin{eqnarray}
  \textbf{v}_\bot^T\textbf{v}_{Q}(\Delta_{\textmd{a}\mid\phi})&=&0,\\
  \lim_{s\rightarrow\infty}\textbf{v}_\bot^T\textbf{v}_{Q}(\Delta_1)&\sim& s^{-2/3}\rightarrow 0.\nonumber
\end{eqnarray}
Furthermore, since generally damping terms do not appear in the dynamic equation for the phase variable, fluctuations in nonlinear damping coefficients are also eliminated at $\Delta_{\textmd{a}\mid\phi}$. For the often encountered case of nonlinear damping proportional to the amplitude cubed~\cite{LCreview,dykmanNonlinearDamping,vanDerPol}, the phase sensitivity at the high saturation limit scales as a constant $(\sim s^0)$ at $\Delta_1$.

Therefore, we have shown that increasing the saturation level of the amplifier and setting the phase-shift to $\Delta_1$, which in this case approaches $\pi$, optimizes the oscillator's performance. This operational mode eliminates the contributions of the feedback noise, thermal noise, and quality factor fluctuations on the oscillator's phase drift.
The phase sensitivity to combined noise sources, whose minimum is obtained for a phase shift value that approaches $\Delta_1$ as saturation grows, and the breakdown to the contributions of the different noise sources, are shown in Fig.~\ref{fig3}. These results were recently verified by experimental phase noise measurements of an oscillator based on a nanomechanical resonator~\cite{guillermo}.

Additional sources of phase noise might originate from the elasticity mechanism of the resonator~\cite{dykman11}. These are expressed by fluctuations in the resonance frequency and nonlinearity coefficient. Fluctuations in the resonance frequency directly translate into phase noise, whose level is approximately constant in the experimental parameters. Fluctuations in the nonlinearity coefficient grow as a function of the saturation level; however, if the fluctuations in both these parameters are correlated, we can potentially design the resonator nonlinearity in such a way so that these contributions cancel each other out.

\section{The oscillator spectrum}\label{spec}

The simplicity of  Eq.~(\ref{proj}) allows us to derive the full oscillator noise spectrum for a variety of noise sources (white, colored, 1/f\ldots). Many of these results may be found in the previous literature \cite{mullen,Demir02,Demir00,vig,lax67,chorti},  but the present approach gives a particularly compact and general formulation for the class of oscillators we consider. We assume a stationary Gaussian noise source with the autocorrelation $\langle\Xi(t_2)\Xi(t_1)\rangle=R_N(t_2-t_1)$, and the spectrum given by a fourier transform $S_N(\omega)={\cal F}[R_N(t)]$. Then $\phi(t+\tau)-\phi(t)=P\int_{t}^{t+\tau}\Xi(t')dt'$ ($\tau>0$) is also Gaussian with  $P=\textbf{v}_\bot^T\textbf{v}_{\textmd{noise}}$ the noise projection constant, and variance $V(\tau) = \langle(\phi(t+\tau)-\phi(t))^{2}\rangle$ given by \cite{Demir02}
\begin{align}\label{var}
V(\tau) & =2P^{2}\int_0^\tau(\tau-v)R_N(v)dv\nonumber\\
& =\frac{4P^{2}}{\pi}\int_{0}^{\infty}S_N(\omega)\left[\frac{\sin(\omega\tau/2)}{\omega}\right]^2d\omega.
\end{align}
The variance $V(\tau)$ typically grows with $\tau$: for a noise source with correlation time $\tau_{c}$ \footnote{We assume a finite correlation time.}, for large enough $\tau\gg\tau_{c}$ the variance increases linearly with time $V(\tau)\simeq P^{2}S_{N}(0)\tau$ corresponding to \emph{phase diffusion}.

The spectrum of the oscillator is the Fourier transform of the autocorrelation function $\langle D(t+\tau)D(t)\rangle$, where $D(t)$ is the displacement of the resonator at time $t$. We take this to be given by $D(t)=a\cos(\omega_{0}t+\phi)$. Neglecting amplitude fluctuations and after transients have died out, the oscillator correlation function is \cite{Demir00,lax66,lax67}
\begin{equation}
\langle D(t+\tau)D(t)\rangle\simeq \frac{a^2}{2}e^{-V(\tau)/2}\cos(\omega_0 \tau)\equiv R(\tau)
\end{equation}
where the Gaussian properties of $\phi$ and the identity $\langle e^{ix}\rangle=e^{-\langle x^{2}\rangle/2}$ for a zero mean Gaussian stochastic variable $x$ have been used.
We see that this autocorrelation function of the oscillator is indeed a stationary stochastic process, as expected for a system without an external time reference.
The oscillator noise spectrum $S(\omega)={\cal F}[R(\tau)]$ can be written as $S(\omega)=a^2[\bar{S}(\omega+\omega_0)+\bar{S}(\omega-\omega_0)]/4$ with
\begin{equation}
\bar{S}(\omega)={\cal F}[e^{-V(\tau)/2}].\label{Fourier}
\end{equation}
The oscillator spectrum about the carrier frequency is thus given by evaluating this Fourier transform.

For large frequency offsets, the Fourier transform in Eq.\ (\ref{Fourier})  is dominated by small times where the variance $V(\tau)$ is small. In this case the exponential can be expanded to first order, so that for $\omega\ne 0$
\begin{equation}
\bar{S}(\omega)\simeq-\tfrac{1}{2}{\cal F}[V(\tau)] =P^{2}\frac{S_{N}(\omega)}{\omega^{2}}.\label{tail}
\end{equation}
This corresponds to the well known Leeson results \cite{Leeson66} for the oscillator noise spectrum, namely an $\omega^{-2}$ dependence for a white noise source, $\omega^{-3}$ for 1/f noise etc. On the other hand, for $\omega\to 0$, the Fourier transform is determined by the long time behavior when the variance $V(\tau)$ gets large, and so the full exponential expression must be used. For small enough noise, $V(\tau)\sim 1$ only at long enough times such that the diffusion behavior $V(\tau)\propto\tau$ applies. In this case the Fourier transform gives a Lorentzian, so that for these frequencies
\begin{equation}\label{lortz}
    \bar{S}(\omega)\simeq\frac{4P^2S_N(0)}{4\omega^2+[P^2S_N(0)]^2}.
\end{equation}
The frequency where the spectrum crosses over from the Lorentzian Eq.\ (\ref{lortz}) to the power law tail Eq.\ (\ref{tail}) will be of order $\bar \tau^{-1}$ where $\bar\tau$ is the time at which the variance $V(\tau)$ grows to be $O(1)$. For a white noise source the Lorentzian leads directly to Eq.\ (\ref{tail}) at large frequencies.

Oscillator phase noise for a general limit cycle and white or colored noise has been looked at previously. For a general limit cycle both the noise vector and projection vector will be time dependent around the limit cycle. Although the oscillator spectrum in the general case is still given by (\ref{Fourier}), the corresponding
 phase variance changes to
\begin{eqnarray}\label{general}
    V_{\textmd{general}}(\tau)
    &=&\int_0^{\tau}du\int_{u-\tau}^{u}dvR_N(v)\\
    &\times&\textbf{v}_\bot^T(u)\textbf{v}_{\textmd{noise}}(u)\textbf{v}_\bot^T(u-v)\textbf{v}_{\textmd{noise}}(u-v).\nonumber
\end{eqnarray}
This much more complicated expression  (cf.\ Eq. (\ref{var})) has only been evaluated in certain limits. Demir et al.~\cite{Demir00} looked at the case of white noise ($\tau_{c}\to 0$) for which expressions Eqs.\ (\ref{tail},\ref{lortz}) apply with the replacement $P^{2}\to t_{p}^{-1}\int_{0}^{t_{p}}(\textbf{v}_\bot^T(t)\textbf{v}_{\textmd{noise}}(t))^{2}dt$ (ie.\ the mean square value of the noise projection). On the other hand Demir \cite{Demir02} looked at the case of colored noise with correlation time much longer than the oscillator period $\tau_{c}\gg t_{p}$, deriving the results Eqs.\ (\ref{tail},\ref{lortz}) but now with the replacement $P\to \int_0^{t_p}\textbf{v}_\bot^T(t)\textbf{v}_{\textmd{noise}}(t)dt$ (ie.\ $P^{2}$ is replace by the square of the mean of the noise projection around the limit cycle). Nakao et al.~\cite{nakao} look at general colored noise, but only derive results for the Lorentzian component of the spectrum  (not the tails further away from the carrier) given by the long time, diffusive behavior of $V(\tau)$. They find a more general expression for the diffusion constant, leading to the replacement in Eq.\ (\ref{lortz})
 $P^{2}S_{N}(0)\to \int_{-\infty}^{\infty}dv R_{N}(s)\,t_{p}^{-1}\int_{0}^{t_{p}}du\textbf{v}_\bot^T(u)\textbf{v}_{\textmd{noise}}(u)\textbf{v}_\bot^T(u-v)\textbf{v}_{\textmd{noise}}(u-v)$. For the special class of oscillators considered in the present work, both $\textbf{v}_{\textmd{noise}},\textbf{v}_\bot^T$ are constant and our expression (\ref{var}) encompasses these limiting cases.

\subsection{Oscillator spectrum for $1/f$ noise}

An interesting special case that arises in may physical implementations is a $1/f$ noise source (or more generally, $1/f^{(\alpha-1)}$ with $1<\alpha\leq2$).
Using the integral
\begin{equation}\label{}
   \int_0^{\infty}\frac{1}{(2 \pi f)^{\alpha}+x^{\alpha}}dx=\frac{2^{1-\alpha}\pi ^{2-\alpha}}{\alpha\sin \left(\frac{\pi }{\alpha}\right)}\left(f^{-\alpha}\right)^{\frac{\alpha-1}{\alpha}},
\end{equation}
a convenient representation of a $1/f^{(\alpha-1)}$ spectrum, with a low frequency cutoff to eliminate the divergence as $f\to 0$ and give a finite correlation time is
\begin{equation}\label{}
S_{N}(f)\propto  \int_{\omega_{c}}^{\infty}\frac{1}{(2 \pi f)^{\alpha}+x^{\alpha}}dx.
\end{equation}
We focus on the pure 1/f case given by $\alpha=2$, so that the noise spectrum is
\begin{eqnarray}
 S_N(f)&=&4I^2\int_{\omega_{c}}^\infty\frac{dx}{x^{2}+(2\pi f)^2}\nonumber\\
 &=&\frac{I^2}{|f|}-\frac{4I^2\arctan(\omega_c/(2\pi f))}{2\pi f}.
\end{eqnarray}
The corresponding phase variance is~\cite{Demir02}
\begin{eqnarray}\label{1overf}
  V(\tau)
   &=&\frac{2 P^2I^2}{\omega_c^2}\bigg[2\omega_c\tau-1+e^{-\omega_c\tau}-\tau\omega_c e^{-\omega_c\tau}\nonumber\\
   &+&(\tau\omega_c)^2E(\tau\omega_c)\bigg],
\end{eqnarray}
with $E(x)=\int_1^{\infty}dye^{-xy}/y$ the exponential integral function. Note that for $\tau\gg\omega_{c}^{-1}$, $V(\tau)\propto \tau$ giving phase diffusion, but for shorter times $\tau \lesssim \omega_{c}^{-1}$
\begin{eqnarray}\label{tau squared log tau}
  V(\tau)&\propto \tau^2 \left(\frac{3}{2}-\gamma-\log (\omega_c\tau)\right),
\end{eqnarray}
where $\gamma\simeq0.58$ is the Euler-Mascheroni constant.

\begin{figure}[]
\begin{center}
      \subfigure[]{
   \includegraphics[width=0.48\columnwidth]{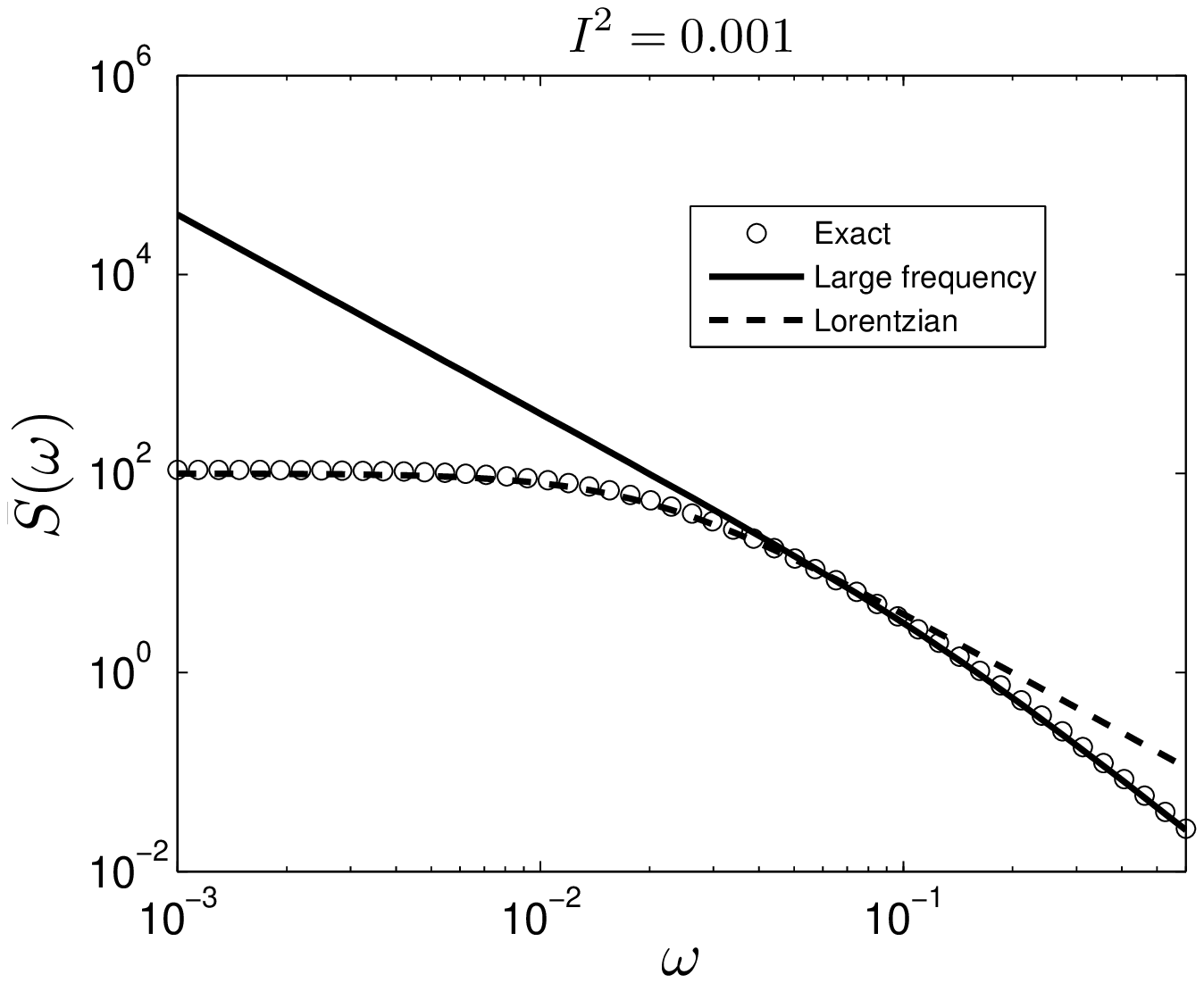}}
  \subfigure[]{
   \includegraphics[width=0.48\columnwidth]{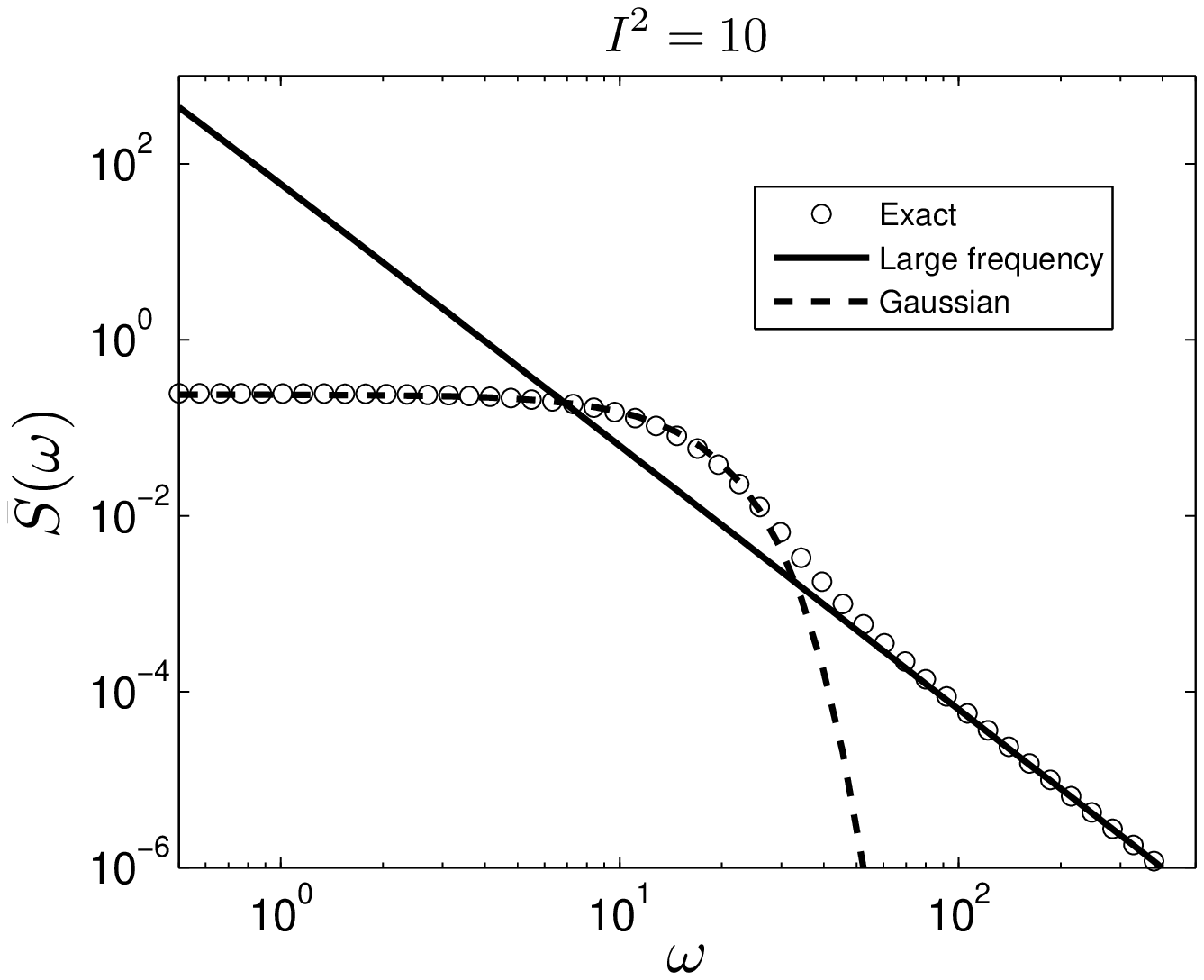}}
   \end{center}
   \caption{\label{fig4} The spectrum of an oscillator subjected to $1/f$ noise. The exact solution is ${\cal F}[e^{-V(\tau)/2}]$ with $V(\tau)$ given by Eq.~(\ref{1overf}). For very weak noise, small offset frequencies are described by the Lorentzian (\ref{lortz}), however for stronger noise, this frequency regime is approximated by the Gaussian (\ref{strong}) as shown in (a) and (b), respectively. Both curves approach the expression (\ref{tail}) at large frequencies. Other parameters are $\omega_c=0.1$, $P^2=1$, and $\omega_m=10$.}
\end{figure}

For very weak noise, the arguments of the preceding section apply, leading to the results Eqs.\ (\ref{lortz},\ref{tail}), ie.\ a Lorentzian with $\omega^{-3}$ tails. However since we might expect the cutoff $\omega_{c}$ to be small, even for not too large noise strengths it is possible for the phase variance $V(\tau)$ to become comparable to unity for times still small enough for Eq. (\ref{tau squared log tau}) to apply rather than the diffusive behavior. The resulting Fourier transform Eq.\ (\ref{Fourier}) can be approximately evaluated in this case by replacing the $\log (\omega_c\tau)$ term in Eq. (\ref{tau squared log tau}) by the constant  $\log(\omega_c/\omega_m)$ where $\omega_m$ represents the frequency range over which we want to approximate. The corresponding spectrum is a \emph{Gaussian}
\begin{equation}\label{strong}
    \bar{S}(\omega)\simeq \frac{\sqrt{\pi}}{2a}e^{-\frac{\omega^2}{4a^2}}
\end{equation}
with $a^2=P^2I^2(3/2-\gamma-\log(\omega_c/\omega_m))$. The resulting oscillator spectra for weak and strong noise intensities are shown in Fig.~\ref{fig4}. A Gaussian spectrum near the carrier frequency also results from the noise spectrum $1/f^{(1-\epsilon)}$ with $0<\epsilon\ll1$~\cite{klimovitch}, and for a rectangle noise spectrum, a Lorentzian and Gaussian are accepted in the weak and strong noise limits, respectively~\cite{stewart}.

\section{Conclusions}

We have demonstrated an analytical method for calculating the phase sensitivity of a class of oscillators, which allows for complete elimination of certain phase noise attributes. We have applied this method to the standard feedback Duffing oscillator, and exposed an operational mode of this oscillator which diminishes both components of the feedback noise and thermal noise, and noise in the damping mechanisms. This therefore offers an optimal operational state of the oscillator which uses the resonator nonlinearity to reduce its phase noise, and disproves the common perception that amplitude-phase noise conversion necessarily degrades the oscillators performance at the nonlinear regime. We have also studied the full oscillator noise spectrum for white and colored noise sources. For a white noise source, or noise with a short correlation time, the near carrier spectrum is Lorentzian, and so varies as $(\omega-\omega_0)^{-2}$ towards larger frequencies. For the often encountered case of $1/f$ noise the near-carrier spectrum is a Lorentzian for weak noise, but as the noise level grows it is well approximated by a Gaussian; in both cases the phase noise spectrum crosses over to $(\omega-\omega_0)^{-3}$ away from the carrier frequency.

This research was supported by DARPA through the DEFYS program.

\appendix

\section{Explicit expressions for special phase shift values}
\subsection{Feedback phase noise}
\label{sec: feedbackApp}
The phase shift values for which feedback phase noise is eliminated are given by equating Eq.~(\ref{ampPhaseNoise}) to zero,
\begin{equation}\label{app1}
    \frac{3 s^2\sin\Delta\cos\Delta}{4}+\frac{1}{2\sin^2\Delta}=0,
\end{equation}
which is equivalent to
\begin{equation}\label{}
    x^4-2x^3+2x+\frac{4}{C}-1=0,
\end{equation}
with $x=\cos2\Delta$, and $C=9s^4/16$. The two real solutions of this equation are

\begin{eqnarray}
  x_1 &=& \frac{1}{2}(1-r+q), \\
  x_2 &=& \frac{1}{2}(1-r-q),\nonumber
\end{eqnarray}
with $r=\sqrt{1+y_R}$, $q=\sqrt{3-r^2+2/r}$, and $y_R$ being a real root of the equation
\begin{equation}\label{cube}
    y^{3}-(1+y)\frac{16}{C}=0,
\end{equation}
whose discriminant is
\begin{equation}\label{}
{\cal D}=-\frac{256 (27 C-64)}{C^3}.
\end{equation}
Since the coefficients of Eq.~(\ref{cube}) are real, for $0<C<64/27$ it has three real roots; however two of them are negative and smaller than -1, so $r$ is imaginary. For the third, positive root, on increasing $C$ the value of $q$ changes from imaginary to real at $C=64/27$, which is the critical Duffing point, corresponding to the drive amplitude $s=(4/3)^{5/4}$~\cite{LCreview}. The corresponding root is
\begin{eqnarray}\label{root}
y_R&=&\frac{2 \left(3^{1/3} \left( \sqrt{3C^3(27 C-64)}+9 C^2\right)^{2/3}+3^{2/3}4C\right)}{3 C \left( \sqrt{3C^3 (27 C-64)}+9 C^2\right)^{1/3}}\nonumber\\
&=&\frac{8}{\sqrt{3C}}\cosh\left[\frac{1}{3}\textmd{arctanh}\left(\frac{\sqrt{3C^3(27C-64)}}{9C^2}\right)\right],
\end{eqnarray}
and the phase shift values which solve Eq.~(\ref{app1}) are
\begin{equation}\label{delta12}
    \Delta_{1,2} = \pi-\frac{\arccos(x_{1,2})}{2}.
\end{equation}
\subsection{Amplitude-phase noise conversion}
\label{sec: ampPhaseApp}
Equation (\ref{ampPhase}) is equivalent to
\begin{equation}\label{cubic}
    (1-x)^3-\frac{1+x}{C}=0,
\end{equation}
whose negative discriminant is
\begin{equation}\label{}
{\cal D}=-\frac{4 (27 C+1)}{C^3}.
\end{equation}
Equation~(\ref{cubic}) thus has one real solution, which is
\begin{eqnarray}\label{}
   x_{\textmd{a}\mid\phi}&=&1+\frac{(\sqrt{3C^3 (27 C+1)}-9 C^2)^{1/3}}{3^{2/3} C}\nonumber\\
   &-&\frac{1}{\sqrt{3}( \sqrt{3C^3 (27 C+1)}-9C^2)^{1/3}},
\end{eqnarray}
and Eq.~(\ref{ampPhase}) is thereby solved by
\begin{equation}\label{DeltaAmpPhase}
    \Delta_{\textmd{a}\mid\phi}=\pi-\frac{\arccos(x_{\textmd{a}\mid\phi})}{2}.
\end{equation}

\bibliography{OptimalOperatingPointReSubmission}

\end{document}